\documentclass{SciPost}

\hypersetup{
    colorlinks,
    linkcolor={red!50!black},
    citecolor={blue!50!black},
    urlcolor={blue!80!black}
}

\usepackage[bitstream-charter]{mathdesign}
\DeclareSymbolFont{usualmathcal}{OMS}{cmsy}{m}{n}
\DeclareSymbolFontAlphabet{\mathcal}{usualmathcal}

\fancypagestyle{SPstyle}{
\fancyhf{}
\lhead{\colorbox{scipostdeepblue}{\bf \color{white} ~SciPost Physics Proceedings }}
\rhead{{\bf \color{scipostdeepblue} ~Submission }}

\fancyfoot[C]{\textbf{\thepage}}
}

\begin{document}

\pagestyle{SPstyle}

\begin{center}{\Large \textbf{\color{scipostdeepblue}{
Graph Neural Network Acceleration on FPGAs for Fast Inference in Future Muon Triggers at HL-LHC\\
}}}\end{center}

\begin{center}\textbf{
Martino Errico\textsuperscript{1,2$\star$},
Davide Fiacco\textsuperscript{1,2},
Stefano Giagu\textsuperscript{1,2},
Giuliano Gustavino\textsuperscript{1,2},
Valerio Ippolito\textsuperscript{2} and
Graziella Russo\textsuperscript{1,2}
}\end{center}

\begin{center}
{\bf 1} Sapienza Università di Roma, Piazzale Aldo Moro 2, Roma, Italy
\\
{\bf 2} INFN Sezione di Roma, Piazzale Aldo Moro 2, Roma, Italy
\\[\baselineskip]
$\star$ \href{mailto:email1}{\small martino.errico@roma1.infn.it}
\end{center}

\definecolor{palegray}{gray}{0.95}
\begin{center}
\colorbox{palegray}{
  \begin{tabular}{rr}
  \begin{minipage}{0.37\textwidth}
    \includegraphics[width=60mm]{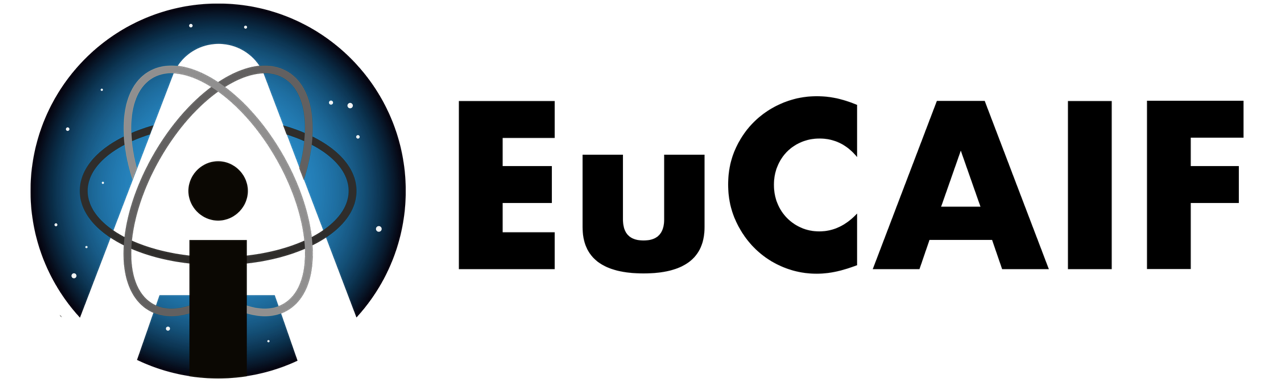}
  \end{minipage}
  &
  \begin{minipage}{0.5\textwidth}
    \vspace{5pt}
    \vspace{0.5\baselineskip} 
    \begin{center} \hspace{5pt}
    {\it The 2nd European AI for Fundamental \\Physics Conference (EuCAIFCon2025)} \\
    {\it Cagliari, Sardinia, 16-20 June 2025
    }
    \vspace{0.5\baselineskip} 
    \vspace{5pt}
    \end{center}
    
  \end{minipage}
\end{tabular}
}
\end{center}

\section*{\color{scipostdeepblue}{Abstract}}
\textbf{\boldmath{
The High-Luminosity LHC (HL-LHC) will reach luminosities up to 7 times higher than the previous run, yielding denser events and larger occupancies. Next generation trigger algorithms must retain reliable selection within a strict latency budget. This work explores machine-learning approaches for future muon triggers, using the ATLAS Muon Spectrometer as a benchmark. A Convolutional Neural Network (CNN) is used as a reference, while a Graph Neural Network (GNN) is introduced as a natural model for sparse detector data. Preliminary single-track studies show that GNNs achieve high efficiency with compact architectures, an encouraging result in view of FPGA deployment.
}}

\vspace{\baselineskip}

\noindent\textcolor{white!90!black}{%
	\fbox{\parbox{0.975\linewidth}{%
			\textcolor{white!40!black}{\begin{tabular}{lr}%
					\begin{minipage}{0.6\textwidth}%
						{\small Copyright attribution to authors. \newline
							This work is a submission to SciPost Phys. Proc. \newline
							License information to appear upon publication. \newline
							Publication information to appear upon publication.}
					\end{minipage} & \begin{minipage}{0.4\textwidth}
						{\small Received Date \newline Accepted Date \newline Published Date}%
					\end{minipage}
			\end{tabular}}
	}}
}

\section{Introduction}
Muons provide clean signatures in collider physics and are essential for many analyses. A muon trigger must keep high efficiency for high-momentum muons while rejecting the muon candidates with momentum under a defined threshold that would otherwise saturate the data acquisition system. The High-Luminosity LHC (HL-LHC) will reach luminosities of about $7.5 \times 10^{34}\,\mathrm{cm^{-2}s^{-1}}$, with pileup levels of up to 200 interactions per bunch crossing~\cite{HL-LHC-TDR}. Under such conditions, conventional hardware-based trigger algorithms face challenges in sustaining sharp thresholds and preserving efficiency. Trigger systems for the HL-LHC are therefore being designed to cope with these demanding requirements while retaining robust performance. Machine-learning techniques offer promising alternatives to conventional trigger algorithms. Convolutional Neural Networks (CNNs) have been studied as image-based classifiers, while Graph Neural Networks (GNNs) naturally capture the sparse and relational structure of detector data and are expected to scale better to complex topologies. In addition to improving efficiency, ML algorithms may also enhance sensitivity to anomalous or unexpected event topologies, which rigid selection rules could fail to capture. Since hardware triggers must operate in real time, Field Programmable Gate Arrays (FPGAs) are a natural choice for deployment. The main challenge is to implement ML models on FPGAs within strict latency ($\mathcal{O}(100\,\mathrm{ns})$) and resource limits. This study explores the viability of real-time ML-based triggering at HL-LHC, presenting the results obtained with a CNN algorithm and the preliminary results produced by a GNN approach when applied to a representative case.

\section{Methodology}
This study adopts an idealized detector inspired by the ATLAS Muon Spectrometer (MS) ~\cite{ATLAS} as a test case, using the geometry expected for its Phase-II upgrade ~\cite{ATLAS-TDR-026}. The barrel of the spectrometer is divided into 32 volumes, 16 sectors of the azimuth for each of the 2 sides of the detector. Each volume includes part of four concentric Resistive Plate Chambers (RPCs) stations at different radial distances (roughly at $r \in \{4.9, 6.8, 7.5, 9.8\} \, \mathrm{m}$) from the beam axis. Each station is independently segmented along pseudorapidity ($\eta$) and azimuth ($\phi$), with strip pitches in the $23 - 35 \, \mathrm{mm}$ range. The spectrometer is immersed in a toroidal magnetic field of roughly $0.5 \, \mathrm{T}$, which bends charged particles in the $(\eta, r)$ plane; thus, the measurements provided by the pseudorapidity strips allow the transverse momentum of muons to be estimated. The RPCs have a time resolution of $1.5 \, \mathrm{ns}$, and their momentum measurements are used for triggering in the barrel region. The first stage (L0) of the Phase-II muon trigger will be implemented on FPGAs with latency constraints of $\mathcal{O}(100\,\mathrm{ns})$, making it a suitable benchmark for ML-based trigger algorithms.

\subsection{Dataset}
The dataset used as a starting point for the training and evaluation of both the CNN and GNN models consists of about $4\times10^{6}$ simulated events, produced accounting for the geometry of the detector. Half contain a muon signal overlaid with background, while the other half are background only. Muons were generated at the interaction point, uniformly in $|\eta| \in (0,1.05)$ and transverse momentum $p_T \in (3,30)\,\mathrm{GeV}$. Background was added according to a simplified model of the LHC cavern background, in which clusters of noise hits, extending along $\eta$, are uniformly distributed across stations and pseudorapidity. The average number of noise clusters per sector per event is 20, reflecting the value expected for HL-LHC. This approximation neglects other sources of noise but reproduces the high occupancies needed to test algorithm performance.

\subsection{CNN approach}
Detector $\eta$-strip hits were binned into $4 \times 384$ arrays, one row per station and one column per $\eta$ bin, then processed with convolutional layers to estimate muon properties. The pseudorapidity binning was chosen to correspond to the typical segmentation in $\eta$-strips of an RPC station, which depends on $r$ in this type of geometry. Event hits were combined to produce multi-muon images, with the dataset being uniformly divided in 0, 1, 2 and 3 muon images. The model predicts the number and momentum of muons present in each image, reconstructing the two leading tracks whenever three are identified. Knowledge Distillation and Quantization-Aware Training were used for model compression~\cite{Francescato2021}.

\subsection{GNN approach}
The $\eta$-strip hits were represented as nodes with $(r,\eta)$ features, and edges were formed between hits in any pair of stations. For different layers, pairs $(i, j)$ were kept only if
\[
|\eta_i - \eta_j| < \frac{|r_i - r_j|}{R_{\min}},
\]
with $R_{\min} = 20 \, \mathrm{m}$ acting as a lower bound for the curvature radius in the magnetic field. Same-layer connections were allowed within a fixed $\eta$ window large enough to account for cluster size. For the GNN dataset, only 0 and 1 muon events were produced. The resulting graphs were processed with an Interaction Network~\cite{Elabd2022}, alternating edge and node updates and ending with global pooling layers for node and edge features to regress the muon transverse momentum.

\subsection{FPGA implementation}
Both models were prepared for FPGA firmware using \texttt{hls4ml}~\cite{Duarte2018,Aarrestad2021}. The CNN contained about 8.4k parameters and synthesized with sub-250\,ns latency on an Ultrascale+ device (XCVU13P). The GNN prototype had about 5.2k parameters; its FPGA implementation is ongoing, with a target latency of $\mathcal{O}(100\,\mathrm{ns})$.

\section{Results}
Efficiency turn-on curves were used as the main performance metric. These curves show, for each true $p_T$ bin, an estimate of the probability that the reconstructed muons pass the given threshold selection, thus illustrating the steep rise towards the efficiency plateau around the cut. They provide a direct way to compare different algorithms in terms of their ability to reject low-$p_T$ muons while maintaining high efficiency for high-$p_T$ muons. 

\begin{figure}[t]
  \centering
  \includegraphics[width=0.8\linewidth]{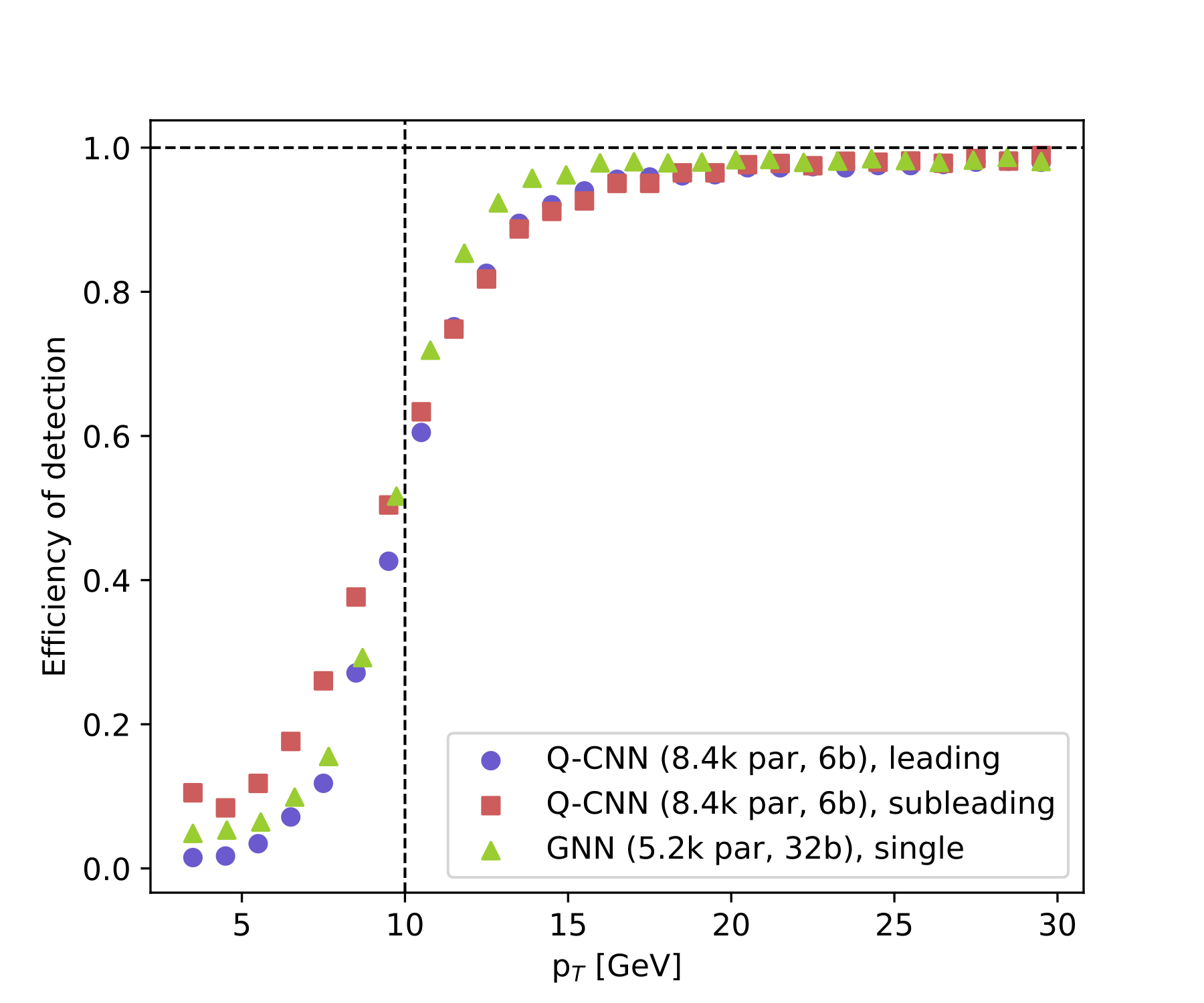}
  \caption{Efficiency turn-on curves: CNN (leading and subleading) vs.\ preliminary GNN results (single-track).}
  \label{fig:eff}
\end{figure}

Figure~\ref{fig:eff} shows the efficiency turn-on curves for the CNN (with muons divided into leading and subleading, where applicable, depending on true momentum) and the GNN (single-track only).

\subsection{CNN performance}
The CNN reached over 90\% efficiency for leading muons with $p_T > 14\,\mathrm{GeV}$ and kept background acceptance below 0.2\%. A clear performance loss was observed only for the subleading track at low momentum, with efficiency around 10\% in the 3--4\,GeV range.

\subsection{GNN performance (single-track)}
The GNN, evaluated on single-track events, achieved efficiencies above 90\% and showed a sharper turn-on than the CNN, indicating better separation of signal from low-$p_T$ muons. These results apply strictly to single-track conditions; however, due to the fact that connectivity information is directly embedded in the graphs and processed by the network, the GNN approach is expected to extend more naturally to multi-muon events than the CNN one.

\section{Conclusions}
The Phase-II ATLAS MS geometry has been employed as a case study to explore machine-learning algorithms for HL-LHC muon triggers. Compressed CNNs confirm the feasibility of image-based inference on FPGAs, but show reduced efficiency for subleading tracks at low $p_T$ due to a low adaptability to overlapping trajectories. A GNN trained and tested on single-muon events achieves competitive efficiency with a smaller number of parameters and is naturally suited to sparse, variable-size inputs, making it a strong candidate for multi-track extension. Hardware implementation of the GNN is in progress, with a target latency of $\mathcal{O}(100\,\mathrm{ns})$.

Future work will focus on the GNN to expand the dataset to multi-muon events, incorporate additional features (e.g.\ $\phi$ and local magnetic field), and complete firmware synthesis and latency validation. Optimization studies (pruning, quantization, ensembles of shallower GNNs) will also be explored. 

These developments represent a promising step toward the integration of modern ML algorithms in next-generation trigger systems.

\section*{Acknowledgements}
This work was partially funded by MUR NextGenEU PRIN2022 I53D23000820006 M4C2.1.1.

\clearpage
\bibliography{bibliography.bib}

\end{document}